\documentclass[10pt]{IEEEtran}
\usepackage{cite}
\usepackage{amsmath,amssymb,amsfonts}
\usepackage{algorithmic}
\usepackage{graphicx}
\usepackage{textcomp}
\usepackage{wrapfig}
\usepackage{booktabs} 
\usepackage{microtype}
\usepackage{graphicx}
\usepackage{subfigure}
\usepackage{booktabs} 
\usepackage{tabularx}
\usepackage{tabulary}

\usepackage{dblfloatfix}
\usepackage{url}
\usepackage{pdflscape}
\usepackage{hyperref}
\usepackage{multirow}
\usepackage{multicol}
\usepackage[english]{babel}
\usepackage{booktabs}
\usepackage{algorithmic}

\usepackage{newfloat}
\DeclareFloatingEnvironment[placement=htp]{algorithm}
\usepackage{caption}

\usepackage{setspace}

\usepackage{comment}
\usepackage{latexsym, amsmath,amssymb, amsthm} 
\newtheorem{thm}{Theorem}
\newtheorem*{thm*}{Theorem}
\newtheorem{dfn}[thm]{Definition}
\usepackage{color}
\usepackage{float}
\usepackage{graphics, epsfig, subfigure, epstopdf, tabularx, footnote, booktabs,pbox, tabularx}
\usepackage{amsfonts}

\usepackage{enumitem}

\newcommand{\hatSr}{\hat{S}_{\rm row}}
\newcommand{\hatSc}{\hat{S}_{\rm column}}

\newcommand{\sE}{\mathcal{E}}
\newcommand{\sL}{\mathcal{L}}
\newcommand{\sU}{\mathcal{U}}

\newcommand{\st}{\;:\;}

\newcommand{\R}{\mathbb{R}}

\begin{document}

\title{Low-Rank Methods in Event Detection \\ and Subsampled Point-to-Subspace Proximity Tests}

\author{Jakub Marecek, Stathis Maroulis, Vana Kalogeraki, and Dimitrios Gunopoulos%
\thanks{Manuscript received \today.}%
\thanks{J. Marecek is at the Czech Technical University, 
Prague, the Czech Republic. 
Email: jakub.marecek@fel.cvut.cz.}
\thanks{S. Maroulis and V. Kalogeraki are at the Athens University 
of Economics and Business, Athens, Greece}
\thanks{D. Gunopoulos is at the University of Athens, Athens, Greece.}
\thanks{This work has received funding from the European Union Horizon 2020 Programme (Horizon2020/2014-2020) under Grant 688380. Jakub has also been supported by OP VVV project CZ.02.1.01/0.0/0.0/16 019/0000765 ``Research Center for Informatics''.}
}

\maketitle

\begin{abstract}
Monitoring of streamed data to detect abnormal behaviour (variously known as event detection, anomaly detection, change detection, or outlier detection) underlies many applications of the Internet of Things. 
There, one often collects data from a variety of sources, with asynchronous sampling, and missing data. In this setting, one can predict abnormal behavior using low-rank techniques. In particular, we assume that normal observations come from a low-rank subspace, prior to being corrupted by a uniformly distributed noise. Correspondingly, we aim to recover a representation of the subspace, and perform event detection by running point-to-subspace distance query for incoming data. 
In particular, we use a variant of low-rank factorisation, which considers interval uncertainty sets around ``known entries'', on a suitable flattening of the input data to obtain a low-rank model. On-line, we compute the distance of incoming data to the low-rank normal subspace and update the subspace to keep it consistent with the seasonal changes present. 
For the distance computation, we suggest to consider subsampling. 
We bound the one-sided error as a function of the number of coordinates employed using techniques from learning theory and computational geometry. 
In our experimental evaluation, we have tested the ability of the proposed algorithm to identify samples of abnormal behavior in induction-loop data from Dublin, Ireland.
\end{abstract}

\

\newcommand{\shortauthors}{J. Marecek et al.}

\maketitle

\section{Introduction}
	
When detailed multivariate data are available in real time, it is highly desirable to monitor the appearance of ``abnormal'' behavior across the multivariate data, with guarantees on the performance of the monitoring procedure, but without the computational burden of processing the data set in its entirety. 
Across the Internet of Things, many examples abound \cite{8369444}. 
To consider one example, many cities have been instrumented with large numbers of sensors capturing 
the numbers and average speeds of cars passing through the approaches of urban intersections (induction loops),  
volume of traffic (from CCTV data or aggregate data of mobile-phone operators),
and speeds of public transport vehicles (e.g., on-board satellite positioning units in buses),
but many still lack the infrastructure to detect traffic accidents prior to them being reported.
This is to a large extent due to the limited utility of the information from each of the sensors,  
\emph{e.g.,} maintaining statistics about traffic at a particular approach of an intersection. 
Only the combination of multivariate time series across multiple sensor types
 could allow the detection of events of interest in many applications.
	
More broadly, there is a monitoring component in most applications of the Internet of Things.
In transportation applications, one may wish to detect traffic accidents \cite{6763098,RSSA:RSSA12178,kong2018lotad,7964673}, 
(imminent) aircraft engine failures \cite{8004441},
or deviations from a flight schedule \cite{clausen2010disruption}.
In electric power distribution systems \cite{farajollahi2017location,7428828,7964673,7908945}, there may be reclosers and sectionalisers acting automatically upon a tree branch falling on an overhead power line,
but the distribution system operator may not know about the event, until it is either detected from sparsely-deployed sensors or reported by customers.
Similar techniques can be used \cite{8930542} in water distribution networks.
Likewise, Internet of Things (IoT) in manufacturing \cite{kanawaday2017machine,9039599} and environmental applications \cite{8081731} crucially relies on monitoring, as does intrusion detection in IoT \cite{8633365,8976157,9199895}, albeit 
the details of the model tend to be more involved and more application-specific. 
Correspondingly, there is a long history of work on monitoring and event detection (also known as anomaly detection or outlier detection), going back at least to \cite{lorden1971procedures,barnett1984outliers} in the univariate case.
Outside of traditional methods, such as dimension reduction \cite{8048463} and Gaussian processes \cite{7428828}, deep-learning methods  \cite{9146846,8794857,9197677,9112195} have been widely used recently. We refer to~\cite{basseville1993detection,8926446} for excellent surveys.

Notice that processing heterogeneous sensor data in IoT applications poses several challenges: 
(1) One of the main challenges is, clearly, dealing with the velocity and, when accumulated, the volume of the data.
A city can have thousands of sensors sampling at kHz rates.  
For example, in a network of $10,000$ sensors, sampling with 1-byte resolution at 1 kHz, one obtains close to 311 TB of data per year 
that needs to be analyzed to estimate what is normal.
(2) The second challenge involves detecting an event in real-time. An automated event detection is useful in cases that the event is 
detected within seconds after it occurs, such as when a road is completely blocked before people start venting their frustration on social media or dialling rescue services. 
(3) Another common challenge is the missing values and failures of sensors. It is widespread for sensors to stop working or start reporting wrong values
 (\emph{e.g.}, negative car flow). Distinguishing the mal-function of a single sensor from a genuine event shows the necessity of utilising multivariate data.  
 (4) Finally, there is measurement noise. In field conditions, e.g., an induction loop buried under 
 inches of tarmac, or a traffic-volume estimate from a video feed captured in a rainstorm, does have a very limited accuracy.
 While there are methods for dealing with each of these challenges in isolation,
 one should like to address all four at the same time.

To overcome these challenges, we propose a novel framework that utilizes low-rank methods \cite{marecek2017matrix} to provide fast and accurate event detection on data 
from varied sources.
Throughout, we consider uniformly-distributed measurement noise, but let us present the model in the noise-free case first in this paragraph.
There, events correspond to points lying outside a certain subspace.
To estimate the sub-space, we flatten the input data to a matrix and apply state-of-the-art low-rank matrix-factorization techniques. 
In particular, we factorize the original matrix into two smaller matrices, whose product approximates the original matrix. 
Subsequently, we develop a point-in-subspace membership test
capable of detecting whether new samples are within the subspace spanned by the columns of one of the factors (smaller matrices).
An affirmative answer is interpreted as an indication that the samples from the sensors present normal behavior.
In the case of a negative answer, a point-to-subspace distance query can estimate the extent of abnormality of an event.
Crucially, this point-in-subspace membership test can be sub-sampled, while still allowing for guarantees on its performance. 
The sub-sampling of, e.g., one per cent of the data, allows for efficient applicability in IoT applications. 


Our main contributions are the following:
     \begin{itemize}
         \item a general framework for representing what is an event and what is a non-event considering \emph{heterogeneous data}, which are possibly \emph{not sampled uniformly}, with \emph{missing values}
        and \emph{measurement errors}.
         \item a novel randomized event detection technique, implemented via a \emph{point-to-subspace distance query}, with guarantees within  \emph{probably approximately correct} (PAC) learning \cite{haussler1990probably}, 
        \item an experimental evaluation on data from a traffic-control system in Dublin, Ireland, which shows that it is possible to process data collected from thousands of sensors over the course of one 
        year within minutes, to answer point-to-subspace distance queries in milliseconds and thus detect even hard-to-detect events.
     \end{itemize}

\section{An Approach}
\label{sec:problem_statement}


Our goal in this paper is to build a model of what is a non-event across many time series,
possibly with \emph{non-uniform sampling} across the time series, \emph{missing values}, 
and \emph{measurement errors} present in the values.
\textcolor{black}{
We build a framework around this model and, in Section~\ref{sec:algos}, suggest algorithms for the individual components in this framework. }

\begin{figure*}[t]
\centering
\includegraphics 
[width=\textwidth]
{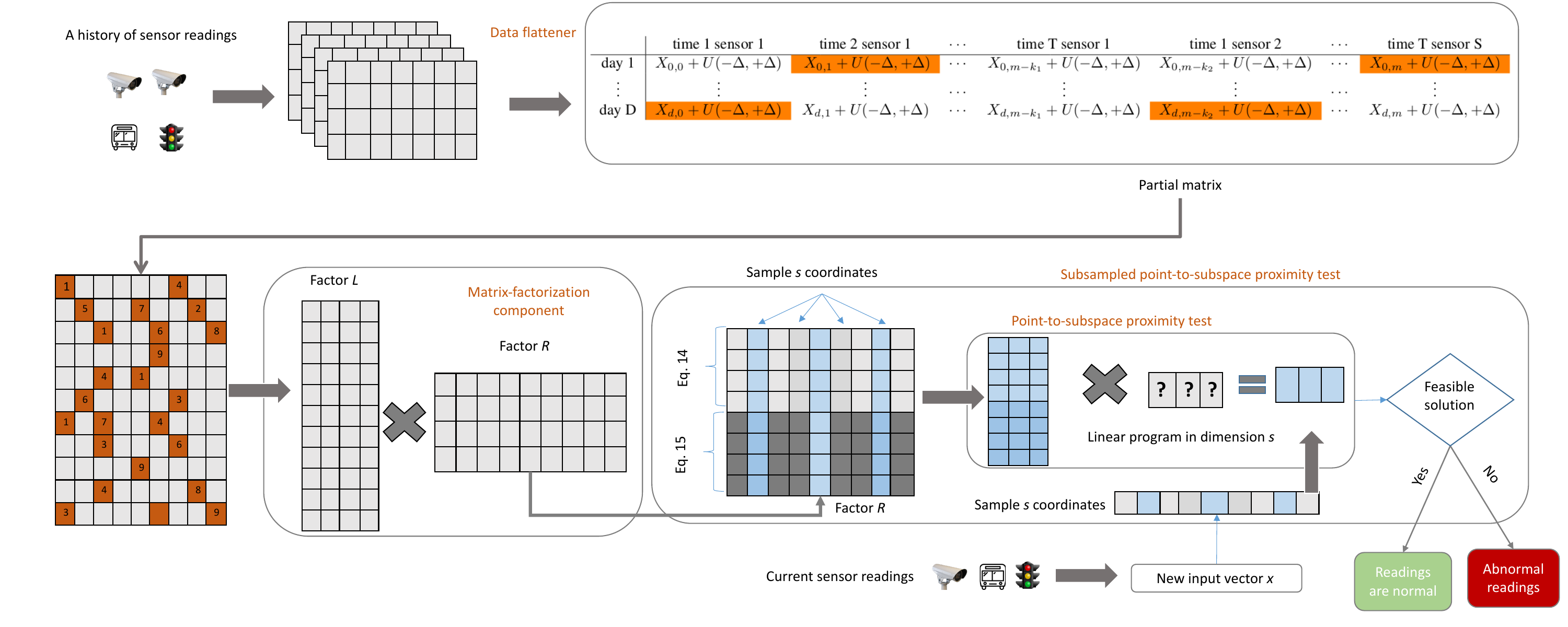}
\caption{A schematic illustration of the MODULoR framework: \textcolor{black}{a history of sensor readings is processed into a partial matrix, which is factorized. One of the factors is subsampled and the corresponding subsampling is applied also to incoming sensor readings. This makes it possible to run a point-to-subspace test in an online fashion, with a constant run time and a small one-sided error}.}
\label{fig:system_model}
\end{figure*}

\textcolor{black}{\subsection{A Model}}
For example, one could consider applications in urban traffic management, where the number of vehicles
passing over induction loops are measured, but often prove to be noisy, with the reliability
of the induction loops and the related communication infrastructure limited.
Subsequently, we aim at an online event detection mechanism, which would be able to decide whether multiple fragments of multiple incoming time-series present an event (abnormal behaviour) or not.
In urban traffic management, for example, one aims at detecting a road accident, based on the evolution
of the traffic volumes across a network of induction loops. Notice that an accident will manifest itself by some readings being low,
due to roads being blocked, while other readings are high, due to re-routing, while no induction loop has to have its readings more than one  standard deviation  away from the long-run average,
which renders univariate methods difficult to use.
Such monitoring problems are central to many Internet-of-Things applications.
This pattern can be exploited by storing each day worth of data as a row in a matrix,
possibly with many missing values.
For multiple time series, we obtain multiple partial matrices, or a partial tensor.
These can be flattened by concatenating the matrices row-wise to obtain one large matrix, as suggested in Figure~\ref{fig:system_model}. 
For $D$ days discretised to $T$ periods each, with up to $S$ sensors available,
the flattened matrix $M$ is in dimension $n = T S$ and has $m = D$ rows.

Considering this flattened representation, it is natural to assume that each new day resembles a linear combination
of $r$ prototypical days, or rows in the flattened matrix in dimension $m \gg r$.
Formally, we assume that there exists $R\in \R^{r \times n}$, such that our observations $x \in \R^n$ are
    \begin{align}
    \label{assumption1}
    x = c R + \mathcal{U}(- \Delta, \Delta),
    \end{align}
    possibly with many values missing,
	for some coefficients $c \in \R^r$ 
    weighing the $r$ vectors $\{ e_1, e_2, \ldots, e_r \}$ row-wise in $R$, with uniformly-distributed noise $\mathcal{U}$
    between $-\Delta$ and $\Delta$.
    
 We compute the matrix $R$, using low-rank approximation of the flattened matrix with an explicit consideration of the uniformly-distributed 
    error in the measurements $M_{ij}$ for $(i,j) \in M$.
    Considering the interval uncertainty set $[M_{ij} -\Delta, M_{ij} + \Delta]$ around each observation, this can be seen 
    as matrix completion with 
    element-wise lower bounds $X^{\sL}_{ij} := M_{ij} -\Delta$ for $(i,j) \in M$ and
    element-wise upper bounds $X^{\sU}_{ij} := M_{ij} + \Delta$ for $(i,j) \in M$.

	%
	%


    Considering the factorization $L R$, where $L \in \R^{m \times r}$ and $R\in \R^{r \times n}$ obtained above \eqref{eq:NONCREF}, 
    given an incoming $x \in \R^n$, the maximum likelihood estimate $\hat c \in \R^r$ of $c$ in \eqref{assumption1} is precisely the point minimizing $|x|_\infty$:
    \begin{align}
    \label{implicitLP}
    \min_{\hat c \in \R^r} \max_i |x_i - (\hat c R)_i| 
    \end{align}
    whenever $\hat c \le \Delta$.
    We refer to Section 7.1.1 of \cite{Boyd2004} for a discussion.
    In a linear program corresponding to \eqref{implicitLP}, we consider a subset of coordinates of $\R^n$ and prove a bound on the one-sided error when using the subset.
    This is the first use of a point-to-subspace query considering the \textcolor{black}{supremum norm ($\ell^\infty$)} in event detection.
       
	

\textcolor{black}{
\subsection{The MODULoR Framework}
This naturally leads to a framework comprising three main components, as illustrated in Figure~\ref{fig:system_model}:
\begin{enumerate}
    \item 
 Data flattener, which captures the unstructured raw input data coming from different sources and reformats them into a partial matrix, which is processed further.
\item 
Matrix-factorization component, which approximates the partial matrix obtained by the data flattener. The approximation consists of two matrices (known as factors), whose product is a low-rank approximation of the original one. Using the two factors, we are able to capture the most salient features of the original matrix in a compressed form. Sparsity of the partial matrix on the input makes the calculation of the matrix factorization possible with high accuracy within modest run time, while allowing for missing values in the input data. This component is further described in Section~\ref{subsec:maco}.
\item 
 \textcolor{black}{Subsampled point-to-subspace proximity tester} (or subspace-proximity tester for short) uses the output of the matrix-factorization component and estimates whether the current sensor readings present an abnormal behavior or not. 
This component is described in detail in Section~\ref{sec:subspace}, but crucially, its run time is independent of the dimension. In experimental results with a history of sensor readings encoded in a partial matrix in $\mathbb{R}^{304 \times 299430}$ and current sensor readings encoded in a vector in $\mathbb{R}^{299430}$, for instance, it takes only milliseconds to perform the test, as we illustrate in of Section~\ref{sec:exp}.
\end{enumerate}
}

The framework can be utilised as follows: Data flattener collects all data from different sources and creates the corresponding matrices, \emph{e.g.}, a partial matrix with traffic volumes and speeds. Then the data structure thus produced is passed to the matrix-factorization component, which factorizes the data, and creates two matrices \emph{L} and \emph{R}. One of the factors (matrix \emph{R}) is then passed to the subspace-proximity tester, which uses it to assess whether incoming sensor readings present abnormal behavior or not and report the results to the end user. Finally, subspace-proximity tester relays the data back to the matrix-factorization component to update the input matrix, replacing the oldest data present, and updating online \cite{feng2013online,guo2014online,akhriev2020pursuit},  if needed.

\textcolor{black}{
We denote this framework MODULoR, where this backronym can stand for 
``MOnitoring Distributed systems Using Low-Rank methods``, or 
more accurately as a ``Method for Outlier Detection Using Low-Rank factorization and range-space subsampling''. We stress that the novelty lies  in the subsampled point-to-subspace proximity tester, whose low run time makes the use of low-rank factorization practical. Without the subsampling, the point-to-subspace distance query in such an approach \cite{feng2013online,6376091,6497613,8847621,zhan2016online,akhriev2020pursuit} would be too demanding for online use. 
}


\section{The Algorithms}
\label{sec:algos}
	
    As outlined above, there are two key algorithms needed. 
    \textcolor{black}{
    The first one implements the matrix-factorization component. In our experiments, we chose the alternating parallel coordinate descent for \emph{inequality-constrained matrix completion} to estimate the low-rank approximation 
    of a partial matrix, either in an online or offline fashion. 
    This makes it possible, rather uniquely, to be robust to  uniformly-distributed measurement noise, while being able to detect sparse noise as abnormal (events, anomalies). 
    }
    
    \textcolor{black}{
	The second algorithm implements the subspace-proximity tester.
	In our experiments, we consider the test with the supremum norm, implemented as a \emph{linear program}, which is subsampled. 
    As an input, it uses the output of the matrix-factorization component
    and it is able to predict if an incoming time series presents normal or abnormal behavior. 
    This second algorithm is run in an online fashion.
    We describe the two algorithms in more detail in the following two sections.
    } 
	
	\subsection{Matrix-factorization Component}
	\label{subsec:maco}
	\begin{algorithm}[t]
		\caption{
		\textcolor{black}{
		Matrix factorization via alternating parallel coordinate descent, cf.  \cite{marecek2017matrix}}}
		\label{alg:SCDM}
		\begin{algorithmic}[1]
			\item[] \textbf{Input}: $\sE, \sL, \sU, X^{\sE}, X^{\sL}, X^{\sU}$, rank $r$
			\item[] \textbf{Output}: $m \times n$ matrix 
			\STATE \label{alg:stp:initialpoint} choose $L\in \R^{m\times r}$ and $R\in \R^{r\times n}$
			\FOR{$k=0,1,2,\dots$}
			\STATE choose a random subset $\hatSr \subset \{1,\dots,m\}$
			\FOR{$i\in \hatSr$ {\bf in parallel}}
			\STATE choose $\hat r \in \{1,\dots,r\}$ uniformly at random
			\STATE compute $\delta_{i \hat r}$ using formula \eqref{eq:delta_L}
			\STATE update $L_{i \hat r}  \leftarrow  L_{i \hat r} + \delta_{i \hat r}$
			\ENDFOR
			\STATE choose a random subset $\hatSc \subset \{1,\dots,n\}$
			\FOR{$j\in \hatSc$ {\bf in parallel}}
			\STATE choose $\hat r \in \{1,\dots,r\}$ uniformly at random
			\STATE compute $\delta_{\hat{r}j}$ using \eqref{eq:delta_R}
			\STATE update $R_{\hat{r}j} \leftarrow R_{\hat{r}j} + \delta_{\hat{r} j}$
			\ENDFOR
			\ENDFOR 
			\STATE \textbf{return} $(L,R)$ 
		\end{algorithmic}
	\end{algorithm}

    To formalise the factorisation $M \approx LR$, 
    let $L_{i:}$ and $R_{:j}$ be the $i$-th row and $j$-th column of $L$ and $R$, respectively.
    With Frobenius-norm regularisation, the factorization problem we wish to solve reads:
	\begin{equation}\label{eq:Specific}
	\min_{L\in \R^{m\times r}, \; R\in \R^{r\times n}} f_{\sL}(L,R) + f_{\sU}(L,R) + \tfrac{\mu}{2}\|L\|_{F}^2 + \tfrac{\mu}{2}\|R\|_{F}^2
	\end{equation}
	where
	\begin{align}\label{defSpecific} 
		f_{\sL}(L,R) &:= \textstyle{\tfrac{1}{2}\sum_{(ij)\in \sL}(X^{\sL}_{ij}-L_{i:}R_{:j})_+^2},\\
		f_{\sU}(L,R) &:= \textstyle{\tfrac{1}{2}\sum_{(ij)\in \sU}(L_{i:}R_{:j}-X^{\sU}_{ij})_+^2},
	\end{align}
	where
    $\xi_+ = \max\{0,\xi\}$,
   calligraphic $\sL$ is used for bounds from below and $\sU$ for bounds from above.
    Notice that this is a non-convex problem, whose special case of $\Delta = 0$ is 
    NP-hard~\cite{MR1320206,harvey2006complexity}.
    
    The matrix completion under interval uncertainty can be seen as a special case of the inequality-constrained matrix completion of \cite{marecek2017matrix}:
	\begin{equation}\label{eq:NONCREF}
	\min \{f(L,R)\;:\; L\in \R^{m\times r}, \; R\in \R^{r\times n}\},
	\end{equation}
	where
	\begin{align}\label{defOff} 
		f(L,R) &:=  f_{\sE}(L,R) + f_{\sL}(L,R) + f_{\sU}(L,R) \\
		       & \; \; + \tfrac{\mu}{2}\|L\|_{F}^2 
		+ \tfrac{\mu}{2}\|R\|_{F}^2 \notag \\
		f_{\sE}(L,R) &:= \textstyle{\tfrac{1}{2}\sum_{(ij)\in \sE}(L_{i:}R_{:j}-X^{\sE}_{ij})^2},\\
		f_{\sL}(L,R) &:= \textstyle{\tfrac{1}{2}\sum_{(ij)\in \sL}(X^{\sL}_{ij}-L_{i:}R_{:j})_+^2},\\
		f_{\sU}(L,R) &:= \textstyle{\tfrac{1}{2}\sum_{(ij)\in \sU}(L_{i:}R_{:j}-X^{\sU}_{ij})_+^2},
	\end{align}
	where
	for $(i,j) \in \sU$ we have an element-wise upper bound $X^{\sU}_{ij}$, 
    for $(i,j) \in \sL$ we have an element-wise lower bound $X^{\sL}_{ij}$,
    for $(i,j) \in \sE$ we know the exact value $X^{\sL}_{ij}$,
    and $\xi_+ = \max\{0,\xi\}$.
    
    \begin{figure*}[t!]
        \centering
        \footnotesize 
    \begin{eqnarray*}  
		\langle \nabla_L f(L,R), E_{i\hat{r}}\rangle = \mu L_{i \hat{r}} + \sum_{j \st (i j) \in \sE}
		( L_{i:} R_{:j} - X^\sE_{ij}) R_{\hat{r}j }  
		+\sum_{j \st (i j) \in \sU  L_{i:} R_{:j} < X_{ij}^\sU }   ( L_{i:} R_{:j}-X_{ij}^\sU) R_{\hat{r}j}
		+\sum_{j \st (i j) \in \sL  L_{i:} R_{:j} > X_{ij}^\sL} (L_{i:} R_{:j}-X_{ij}^\sL) R_{\hat{r}j}\\
		\langle \nabla_R f(L,R), E_{\hat{r}j}\rangle = 
		\mu R_{\hat{r}j} + \sum_{i \st (ij) \in \sE}
		( L_{i:} R_{:j} - X^\sE_{ij}) L_{i\hat{r}} 
		+ \sum_{i \st (ij) \in \sL  L_{i:} R_{:j} < X_{ij}^\sL }
		( L_{i:} R_{:j}-X_{ij}^\sL) L_{i\hat{r}} 
		+ \sum_{i \st (ij) \in \sU  L_{i:} R_{:j} > X_{ij}^\sU }
		(L_{i:} R_{:j}-X_{ij}^\sU) L_{i\hat{r}}.
	\end{eqnarray*}
	\caption{Multiplication by the gradients in \eqref{eq:delta_L} and  \eqref{eq:delta_R} of Algorithm~\ref{alg:SCDM} can be simplified considerably.}
     \label{fig:gradients}
        \end{figure*}

	A popular heuristic for matrix completion considers a product of
	two matrices, $X=L R$, where $L \in \R^{m \times r}$ and $R\in \R^{r \times n}$, obtaining 
    $X = LR$ of rank at most $r$, cf. \cite{tanner2013normalized}.
	In particular, we use a variant of the alternating parallel coordinate descent method for matrix completion introduced by 
	\cite{marecek2017matrix} under the name of ``MACO'',
	summarized in Algorithm~\ref{alg:SCDM}.
	It is based on the observation that while $f$ is not convex jointly
	in $(L,R)$, it is convex in $L$ for fixed $R$ and in $L$ for fixed $R$.
	In Steps 3--8 of the algorithm, we fix $R$, choose random $\hat{r}$ and a random set $\hatSr$ of rows of $L$, and
	update, in parallel, for $i \in \hatSr$:  $L_{i\hat{r}} \leftarrow L_{i\hat{r}} + \delta_{i\hat{r}}$. Following~\cite{marecek2017matrix}, we use
	\begin{equation} \label{eq:delta_L}\delta_{i\hat{r}}:= - \langle \nabla_L f(L,R), E_{i\hat{r}}\rangle / W_{i\hat{r}},\end{equation}
	where the computation of $\langle \nabla_L f(L,R), E_{\hat{r}j}\rangle$ can be simplified as suggested in Figure \ref{fig:gradients}.
	In Steps 9--14, we fix $L$, choose random $\hat{r}$ and a random set $\hatSc$ of columns of $R$, and update, in parallel
	for $j \in \hatSc$: $R_{\hat{r}j}\leftarrow R_{\hat{r}j} + \delta_{\hat{r}j}$.
	\begin{equation} \label{eq:delta_R}\delta_{\hat{r}j}:= - \langle \nabla_R f(L,R), E_{\hat{r}j}\rangle / V_{\hat{r}j},\end{equation}
	where the computation of $\langle \nabla_R f(L,R), E_{\hat{r}j}\rangle$ can, again, be simplified as suggested in Figure \ref{fig:gradients}.
	
    
	We should also like to comment on the choice of $\Delta$ and $\epsilon$. 
    A sensible approach seems to be based on cross-validation: out of the historical data (or out of $L$), 
    one can pick one row, and compute the $\Delta$ needed. The maximum of $\Delta$ for any row seems to be a good choice.
	We refer to~\cite{marecek2017matrix} for a discussion
	of the choice of the parameter $\mu>0$.        
	
\subsection{Subsampled Point-to-Subspace Proximity Tester}
\label{sec:subspace}
	
    As suggested previously, instead of computing the distance of an incoming time-series to each one of those already available per-day time-series, classified as event or non-event, 
    we consider a point-to-subspace query in the infinity norm:
    \begin{align}
    \label{eq:infty}
    \min_{\hat c \in \R^r} \max_i |x_i - (\hat c R)_i|, 
    \end{align}
    and test whether the distance \eqref{eq:infty} is less than or equal to $\Delta$.
    As we described in Section~\ref{sec:problem_statement} for uniform noise\textcolor{black}{, the supremum norm ($\ell^\infty$)} gives the maximum likelihood estimate. The infinity norm is sometimes seen as difficult to work with, due of the lack of differentiability.
    However, note that it  \eqref{eq:infty} can be recast as a test of the feasibility of a linear programming problem:
    \begin{align}
    \label{eq:inftyLP}
    \min_{\hat c \in \R^r} 1 \; \textrm{s.t. }  x_i - (\hat c R)_i & \le \Delta, \\
                                                (\hat c R)_i - x_i & \le \Delta. \label{eq:inftyLP2}
    \end{align}    
    Alternatively, this is an intersection of  hyperplanes, also known as a hyper-plane arrangement. As we will show in the following section, this geometric intuition is useful in the analysis of the algorithms.

    In Algorithm \ref{alg:SPC} we present a test, which considers only a subset $S, |S| \ll n$ of coordinates, picked uniformly at random.
    As we show in the following section,
    this test has only a modest one-sided error.

	\section{An Analysis}
	\label{sec:anal}
	
	Before we present the main result, let us remark on the convergence properties of  Algorithm \ref{alg:SCDM},
	which has been proposed and analyzed by \cite{marecek2017matrix} and \cite{akhriev2020pursuit}. 
    A simple convergence result of \cite{marecek2017matrix} states that the method is monotonic and, with probability 1,
    $\lim_{k\to \infty} \inf \|\nabla_L f(L^{(k)},R^{(k)})\|
    = 0,$ 
    and  
    $\lim_{k\to \infty} \inf \|\nabla_R 
    f(L^{(k)},R^{(k)})\| = 0.$
    This applies in our case as well.
    See \cite{akhriev2020pursuit} for details of the rate of convergence. 

    Our main analytical result concerns the statistical performance  
	of the point-to-subspace query.
    Informally, 
    the randomized point-to-subspace distance query in Algorithm \ref{alg:SPC}
    has one-sided error: 
    If the distance between the vector $x$ and $\textrm{span}(R)$
    is no more than  $\Delta$ in $\ell^\infty$, we never report otherwise.
    If, however, the distance actually is more than $\Delta$ in $\ell^\infty$,
    considering only a subset $S$ of coordinates may ignore a coordinate where the distance is larger,
    and hence mis-report that the vector is within distance $\Delta$ in $\ell^\infty$,
    with a certain probability, depending on the number of constraints that
    are actually violated.
    For example, to achieve the one-sided error of $\epsilon$ with probability of 1/3 or less, this test needs to solve a linear program in dimension 
    $O(\frac{r \log r}{\epsilon} \log\frac{r \log r}{\epsilon})$. 
    Notice that this bound is independent of the ``ambient'' dimension $n$.
    
    Formally:
    
    



    \begin{thm}
    \label{ourThm}
    (i) When the distance \eqref{eq:infty} is $D \le \Delta$, Algorithm \ref{alg:SPC} never reports the point is outside the sub-space.
    (ii) When the distance \eqref{eq:infty} is $D > \Delta$,
    because there are $\epsilon n$ coordinates $i$ such that for all $\hat c$, there is $|x_i - (\hat c R)_i| \ge \Delta$,
    then for any $\delta \in (0, 1)$,
    when Algorithm \ref{alg:SPC} 
    considers $s$ coordinates 
    $$O\left( \frac{1}{\epsilon} \log \frac{1}{\delta} + \frac{r \log r}{\epsilon} \log \frac{r \log r}{\epsilon} \right)$$
    sampled independently uniformly at random, the point is inside the subspace with probability $1 - \delta$.
    \end{thm}
    

	\begin{algorithm}[t!]
		\caption{\textcolor{black}{Subsampled point-to-subspace proximity tester with the supremum norm}
		}
		\label{alg:SPC}
		\begin{algorithmic}[1]
			\item[] \textbf{Input}: $R \in \R^{r \times n}$,
			$x \in \R^n$,
			$s, \Delta \in \R$
			\item[] \textbf{Output}: true/false
			\STATE 
			\label{line:sample}
			choose $S \subset \{1,\dots,n\}, |S| = s$,  uniformly at random
			\STATE initialise a linear program $P$ in variable $v \in \R^s$
			\label{line:init}
			\FOR{$i\in S$ } 
			\STATE 
			\label{constraint1}
			add constraint $x_i - (\textrm{proj}_S(L) v)_i \le \Delta$
			\STATE 
			\label{constraint2}
			add constraint $x_i - (\textrm{proj}_S(L) v)_i \ge - \Delta$
			\ENDFOR
			\IF{$\exists v \in \R^s$ such that the constraints are satisfied}
			\label{line:solveLP}
			\STATE \textbf{return} true
			\ELSE
			\STATE \textbf{return} false
			\ENDIF
		\end{algorithmic}
	\end{algorithm}
    \begin{proof} \emph{(Sketch)} \;
    To see (i), consider the linear program constructed in Algorithm \ref{alg:SPC} and 
    notice that its constraints are a subset of those in \eqref{eq:inftyLP}.
    If \eqref{eq:inftyLP} is feasible, then any subset of constraints will be feasible.
    To see (ii), 
    we use standard tools from computational geometry.
    In particular, we show that 
     a certain set related to the polyhedron of feasible $x$, which is known as range space, has a small Vapnik-Chervonenkis (VC) dimension $d$. 
    Subsequently, we apply the celebrated result of \cite{Haussler1987}, which states that for any range space of VC dimension $d$ and $\epsilon, \delta \in (0, 1)$, if $$O\left( \frac{1}{\epsilon} \log \frac{1}{\delta} + \frac{d}{\epsilon} \log \frac{d}{\epsilon} \right)$$ 
    coordinates are sampled independently, we obtain an $\epsilon$-net with probability at least $1 - \delta$.
     We refer to Appendix A for details.
    \end{proof}
    
    

    Next, let us consider the run-time of 
    Algorithm \ref{alg:SPC}, which is dominated by the feasibility test of a linear program $P$ in Line \ref{line:solveLP}.
    Using standard interior-point methods \cite{GONDZIO2012},
    if there is a feasible solution to the linear program $P$, an $\epsilon$-accurate approximation to the 
    can be obtained in $O(\sqrt{s} \ln(1/\epsilon))$
    iterations, wherein each iteration
    amounts to solving a linear system. 
    This yields an upper bound on the run-time of
   $$O \left( \frac{r^{3.5} \log^{3.5} r}{\epsilon^{3.5}} \log^{3.5}\frac{r}{\epsilon} \right),$$
    which could be improved considerably by exploiting the sparsity in the 
    linear program's constraint matrix.
    The same iterations make it possible to detect
    infeasibility using the arguments of \cite{kojima1993general},
    although the homogeneous self-dual approach of \cite{Ye1994}
    with a  worse iteration complexity may be preferable in practice.
    Either way, a solver generator \cite{mattingley2010real,mattingley2012cvxgen} allows for excellent performance. 

    Alternatively, however, one may consider:

    \begin{thm}
    \label{efficient}
    There is an algorithm 
    that can pre-process a sample of $s$ coordinates such that the point-in-subspace membership query can be answered in time $O(\log s)$ in the worst case.
    The expected run-time of the pre-processing is  $O(s^{r+\epsilon}), \epsilon \ge 0$, where the expectation is with respect to the random behaviour of the algorithm, and remains valid for any input. 
    \end{thm}
    
    \begin{proof} \emph{(Sketch)} \;
    Notice that one can replace the test of feasibility of a linear program $P$ with a point-location problem in a hyperplane arrangement.
    We refer to \cite{deBerg2000,stanley2004} for a very good introduction to hyperplane arrangements, but to provide an elementary intuition:     
    An alternative geometric view of Algorithm \ref{alg:SPC} is that we have a subspace $P \subseteq \R^s$,
    initialise $P = \R^s$ in Line \ref{line:init}, 
    and then intersect it with hyperplanes on Lines \ref{constraint1}--\ref{constraint2}.
    Equally well, one may consider a hyper-plane arrangement $P$,
    initialise it to an empty set in Line \ref{line:init}, 
    and then add hyperplanes on Lines \ref{constraint1}--\ref{constraint2}.
    Our goal is not to optimise a linear function over $P$, but rather to decide whether there exists a point within $P$, the intersection of the hyperplanes, which corresponds to one cell of the arrangement.
    The actual result follows from the work of \cite{Clarkson1986,Clarkson1987} on hyperplane arrangements.
    \end{proof}

    While the use of solver-generator \cite{mattingley2010real,mattingley2012cvxgen} may be preferrable in many  IoT applications, there may be large-scale use cases, where the asymptote of the run-time of the algorithm of \cite{Clarkson1987} does matter and the sampling of the coordinates may be reused.
\begin{figure*}[!t]
\centering
\subfigure[]{%
		\includegraphics*[width=0.31\textwidth]{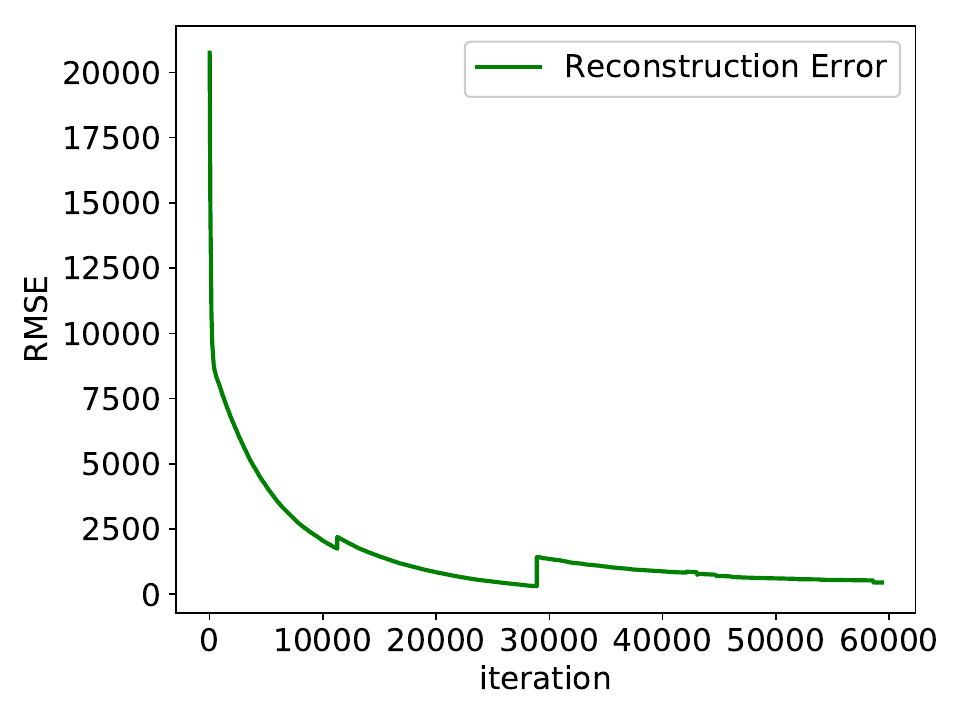}
        \label{fig:iterations}}
        	\def\subfigcapskip{-0.1cm}
	\subfigure[]{%
		\includegraphics*[width=0.31\textwidth]{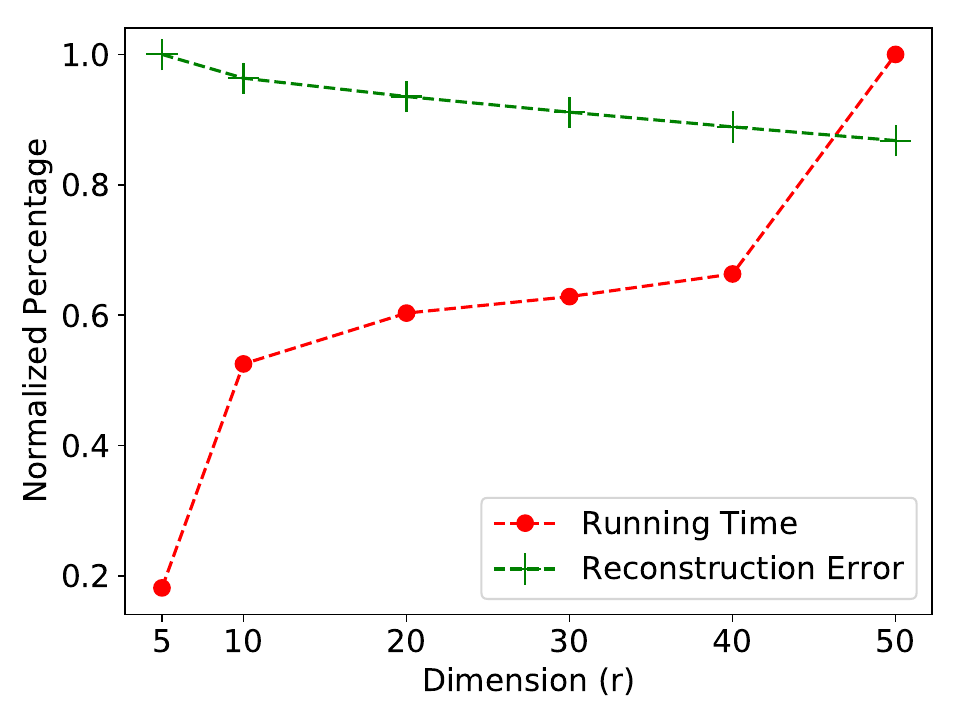}
        \label{fig:time_error_comparison}}
    \subfigure[]{
        	\includegraphics*[width=0.31\textwidth]{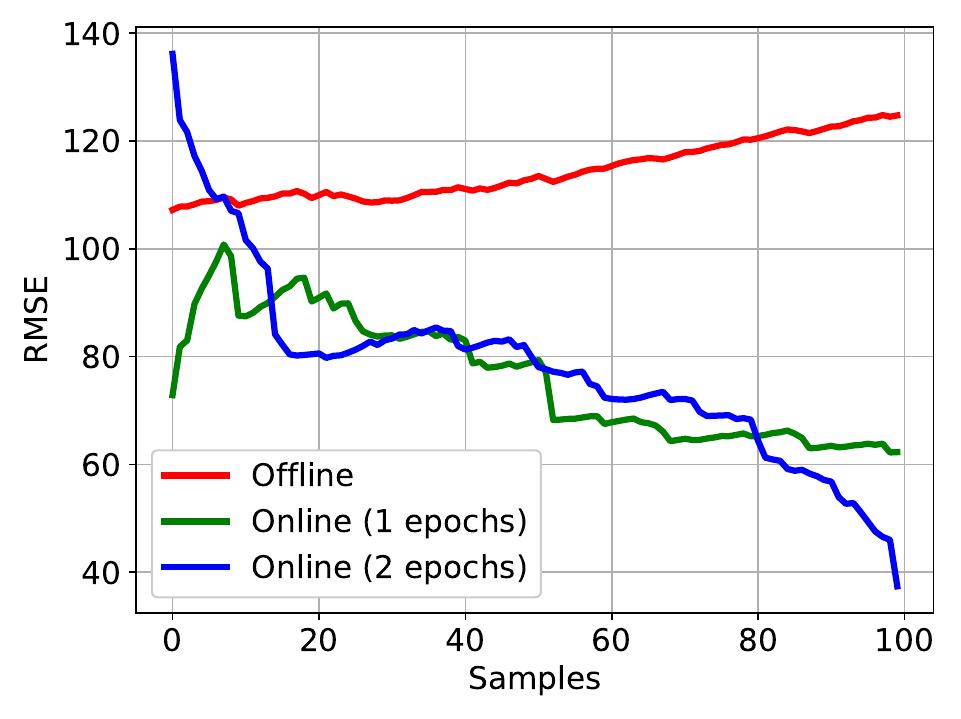}
        	 \label{fig:retrain}}
        \centering
		\caption{
		\textcolor{black}{Performance of the matrix-factorization component on the instance of Section \ref{sec:data}.  
		Figure~\ref{fig:iterations}
		presents one sample evolution of the reconstruction error over time for $r =10$.
		Figure~\ref{fig:time_error_comparison}
		displays the reconstruction error and training time (until improvement in the error falls below $10^{-4}$), both as  functions of rank $r$.
		Notice that the approach seems rather robust to the choice of $r$.
		Figure~\ref{fig:time_error_comparison_iterations}
		compares the evolution of reconstruction error for three variants of the method, as described in the text.
		Notice that online variants seem superior to the offline variant.\\[6mm]
		}}
\label{fig:time_error_comparison_iterations}
\end{figure*}

\begin{figure*}[!pt]
\centering
		\subfigure[]{%
			\label{fig:deviation}
			\includegraphics*[width=0.31\textwidth]{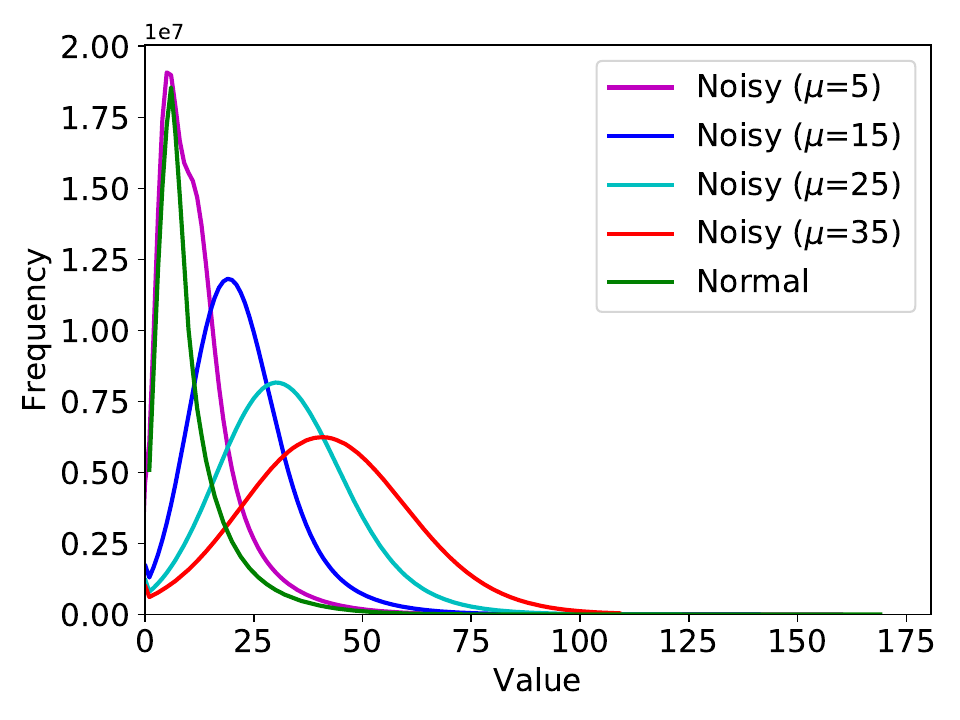}}
		\def\subfigcapskip{-0.1cm}
		\subfigure[]{%
		\includegraphics*[width=0.31\textwidth]{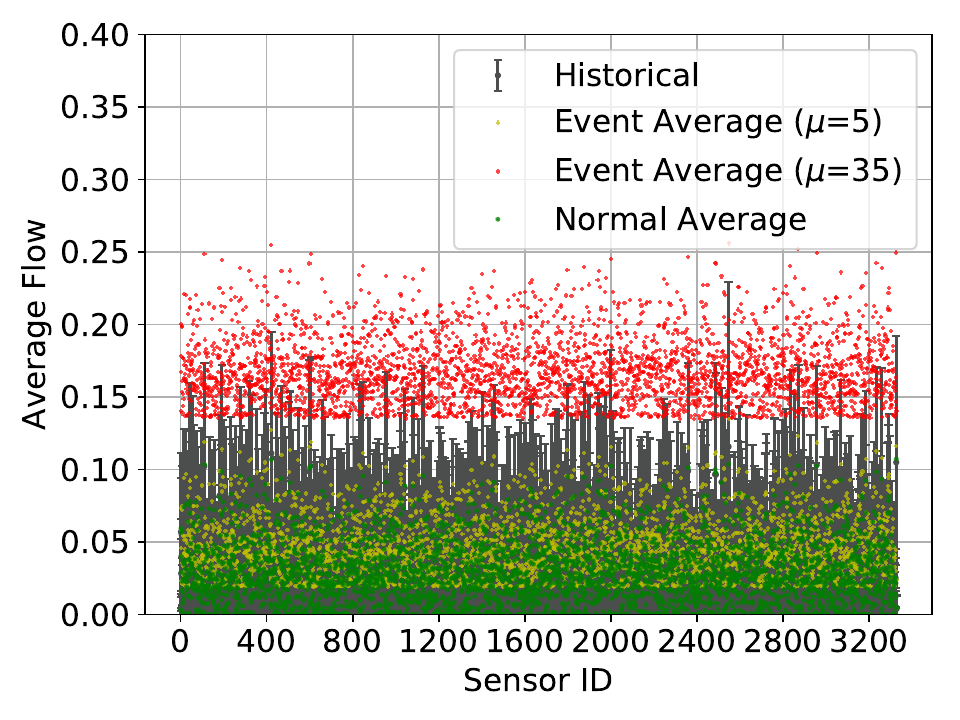}
        \label{fig:distance_historical}}
        	\def\subfigcapskip{-0.1cm}
		\subfigure[]{%
		\includegraphics*[width=0.31\textwidth]{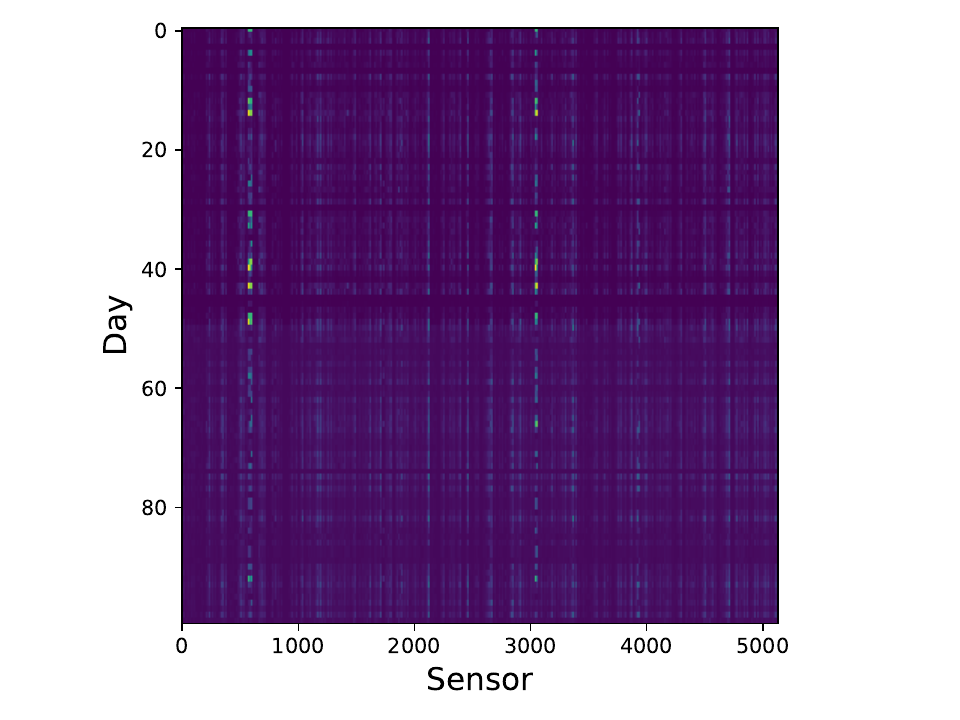}
        \label{fig:heatmap}}
        	\def\subfigcapskip{-0.1cm}
		\caption{
		\textcolor{black}{
		Properties of the instance of Section \ref{sec:data}:
		Figure~\ref{fig:deviation} shows frequencies of the values reported from sensors, with and without additional Gaussian noise. Figure~\ref{fig:distance_historical} illustrates the mean and standard deviation of the historical flow data at all available sensors (grey), plus the mean values for events (yellow for $\mu = 5$, red for $\mu =35$) and non-events (green) at the same sensors. Finally, in Figure~\ref{fig:heatmap}, there is a heatmap of a validation matrix, which contains the normal reading (upper half) and event readings with $\mu =35$ (lower half).}\\[6mm]}
\end{figure*}

\begin{figure*}[!pt]
		\centering
		\def\subfigcapskip{-0.1cm}
		\subfigure[Recall]{%
			\label{fig:recall_synth}
			\includegraphics*[width=0.31\textwidth]{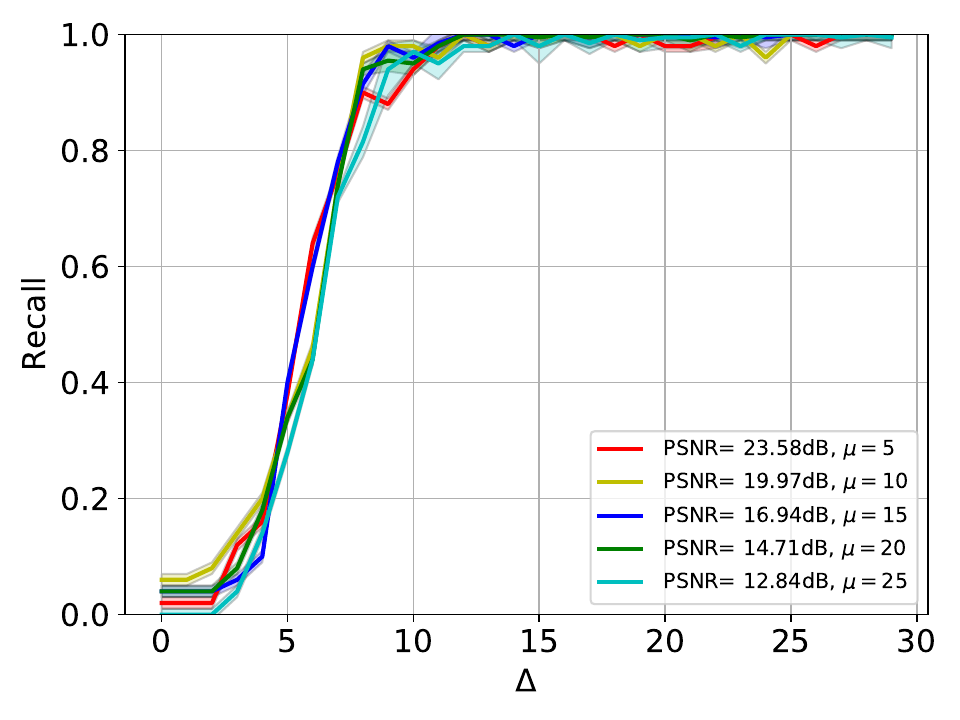}}
		\def\subfigcapskip{-0.1cm}
		\subfigure[Precision]{%
			\label{fig:precision_synth}
			\includegraphics[width = 0.31\textwidth]{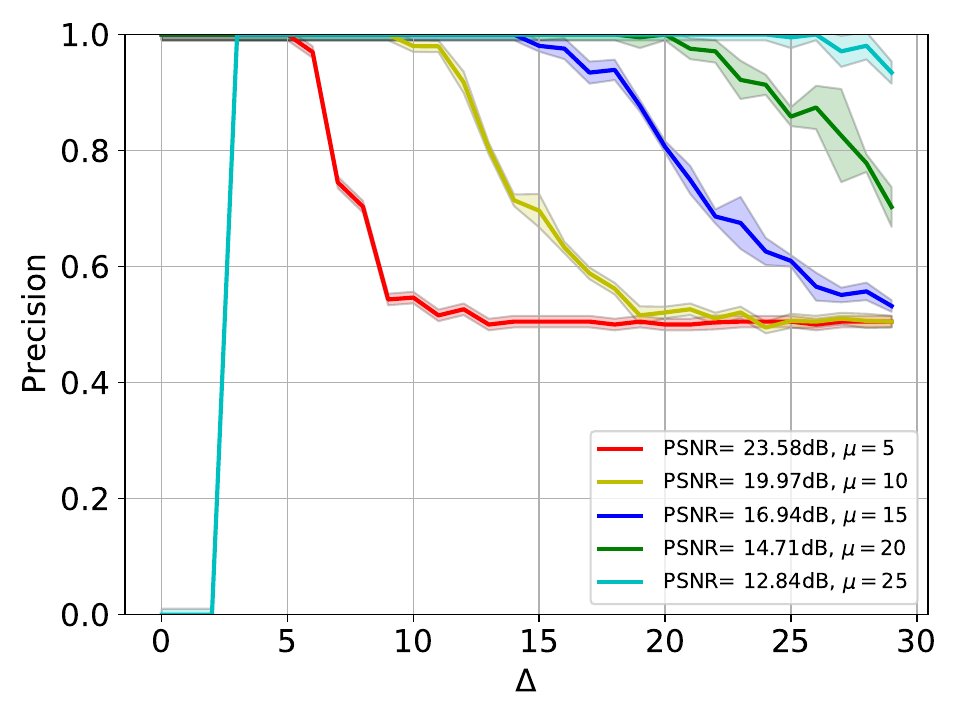}}
		\def\subfigcapskip{-0.1cm}
		\subfigure[F1-Score]{%
			\label{fig:f1_synth}
			\includegraphics*[width=0.31\textwidth]{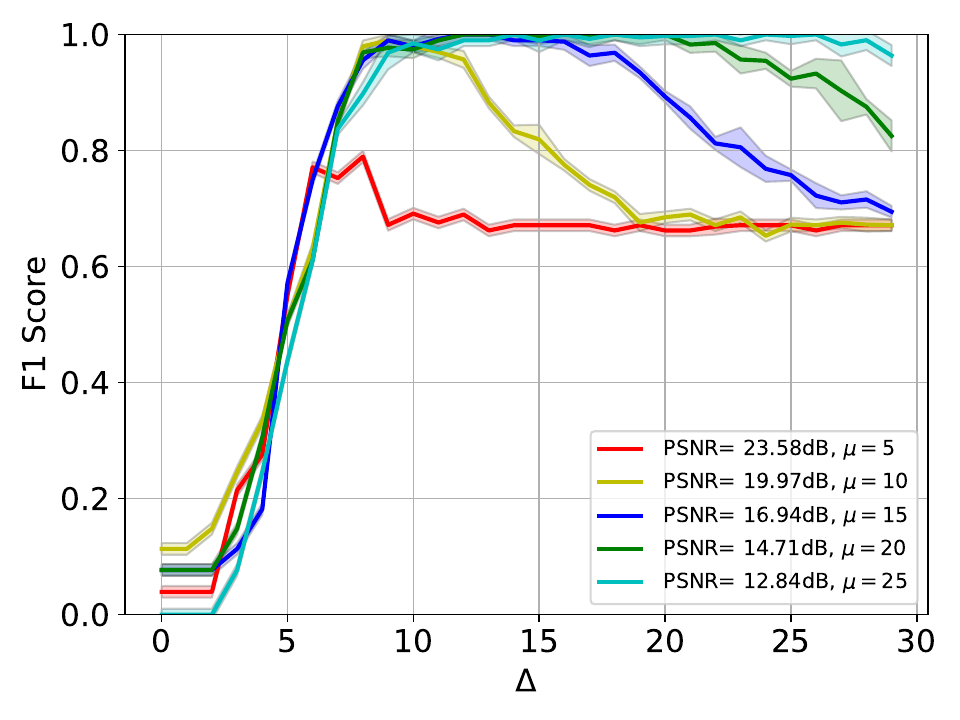}}
		\def\subfigcapskip{-0.1cm}
		
		\caption{\textcolor{black}{Results of repeated six-fold cross validation as a function of the half-width $\Delta$ of the uncertainty set and the PSNR used to generate the synthetic instance of Section \ref{sec:data}: Mean (solid line) and one standard deviation (half-width of the semi-transparent error band around the solid line) of three performance measures (recall, precision and F1 score). 
		}\\[6mm]}
		\label{fig:syth_recall_precision_f1}
\end{figure*}

\begin{figure*}[!t]
		\centering
		\def\subfigcapskip{-0.1cm}
		\subfigure[Recall]{%
			\label{fig:var_recall_synth}
			\includegraphics*[width=0.3\textwidth]{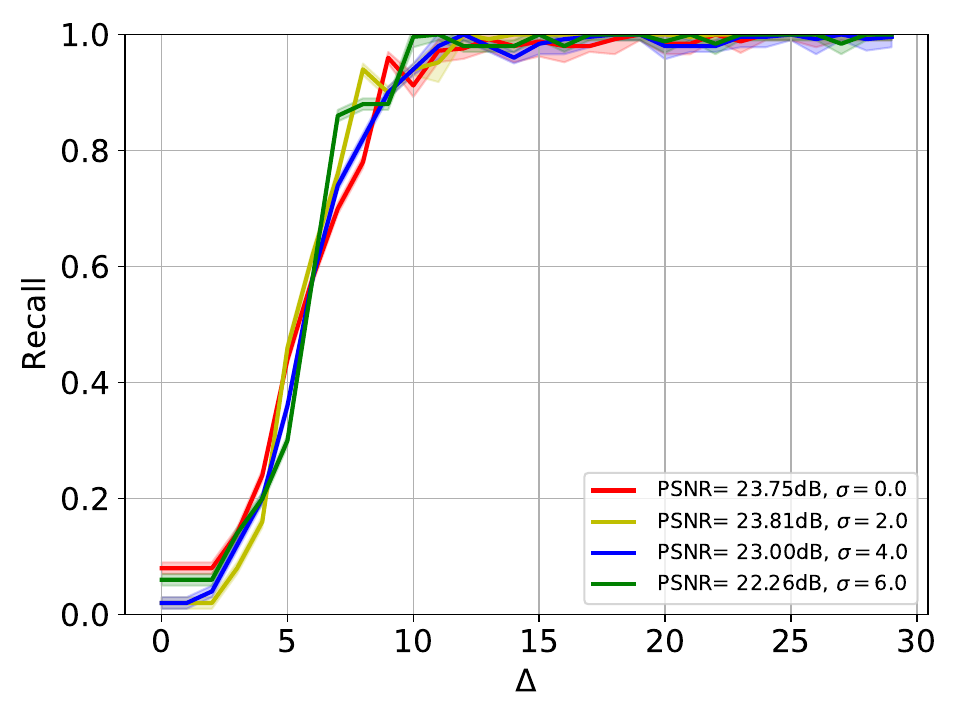}}
		\def\subfigcapskip{-0.1cm}
		\subfigure[Precision]{%
			\label{fig:var_precision_synth}
			\includegraphics[width = 0.3\textwidth]{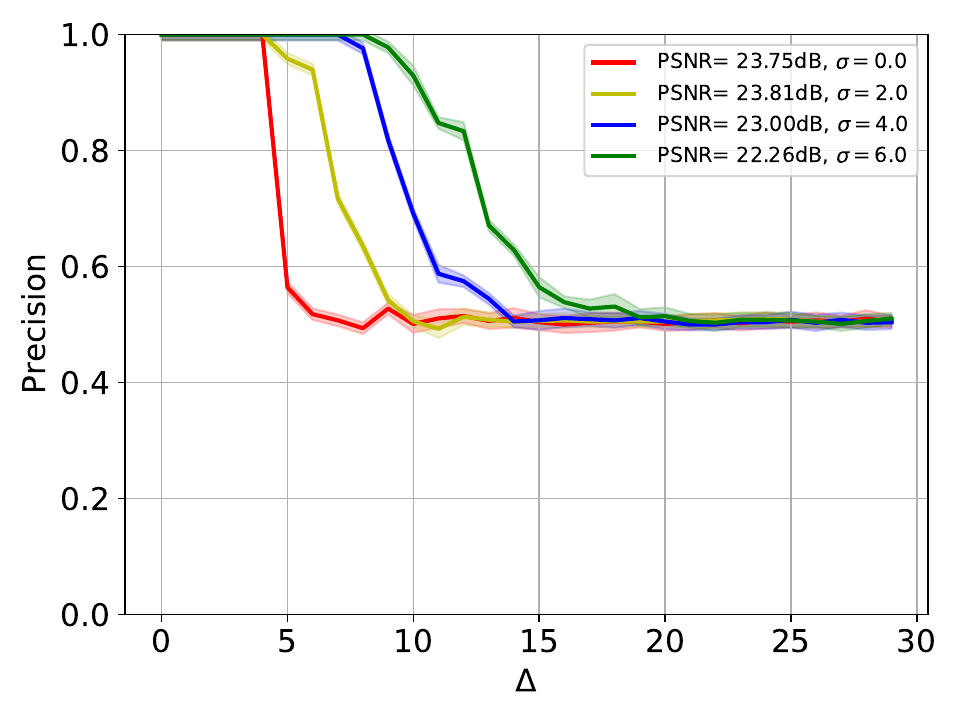}}
		\def\subfigcapskip{-0.1cm}
		\subfigure[F1-Score]{%
			\label{fig:var_f1_synth}
			\includegraphics*[width=0.3\textwidth]{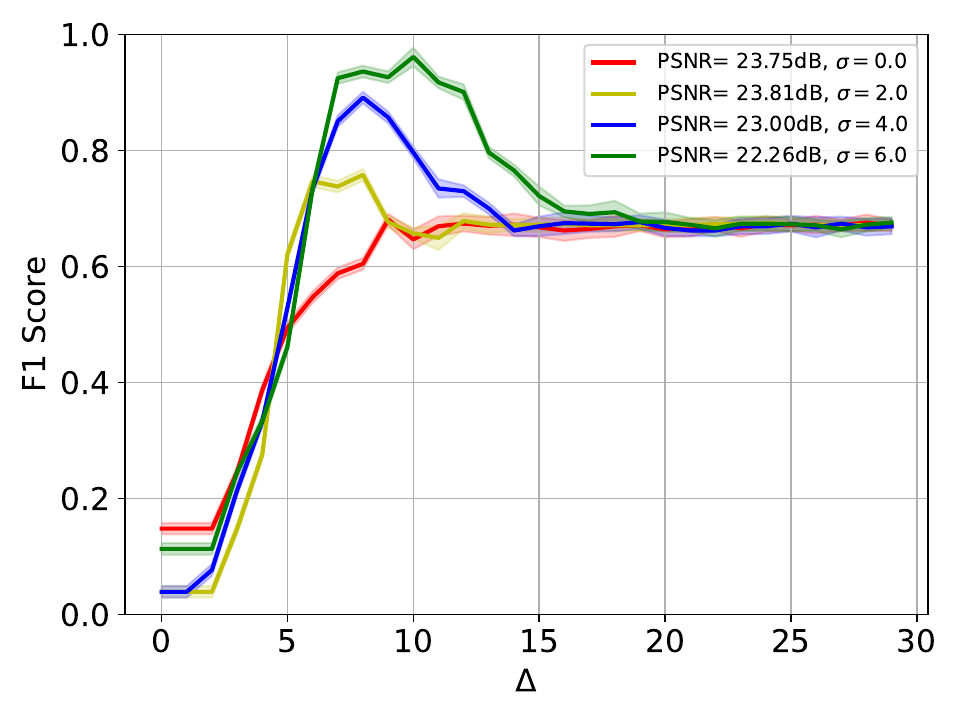}}
		\def\subfigcapskip{-0.1cm}
		
		\caption{\textcolor{black}{Results of repeated six-fold cross validation as a function of the half-width $\Delta$ of the uncertainty set and the PSNR used to generate the synthetic instance of Section \ref{sec:data}: Mean (solid line) and one standard deviation (half-width of the semi-transparent error band around the solid line) of three performance measures (recall, precision and F1 score). 
		The variation of PSNR is solely due to varying the standard deviation of the noise used to generate the synthetic instance, while keeping its mean low and constant at $\mu = 5$.}
		}
		\label{fig:var_syth_recall_precision_f1}
\end{figure*}

\begin{figure}[!t]
			\includegraphics*[width=\columnwidth]{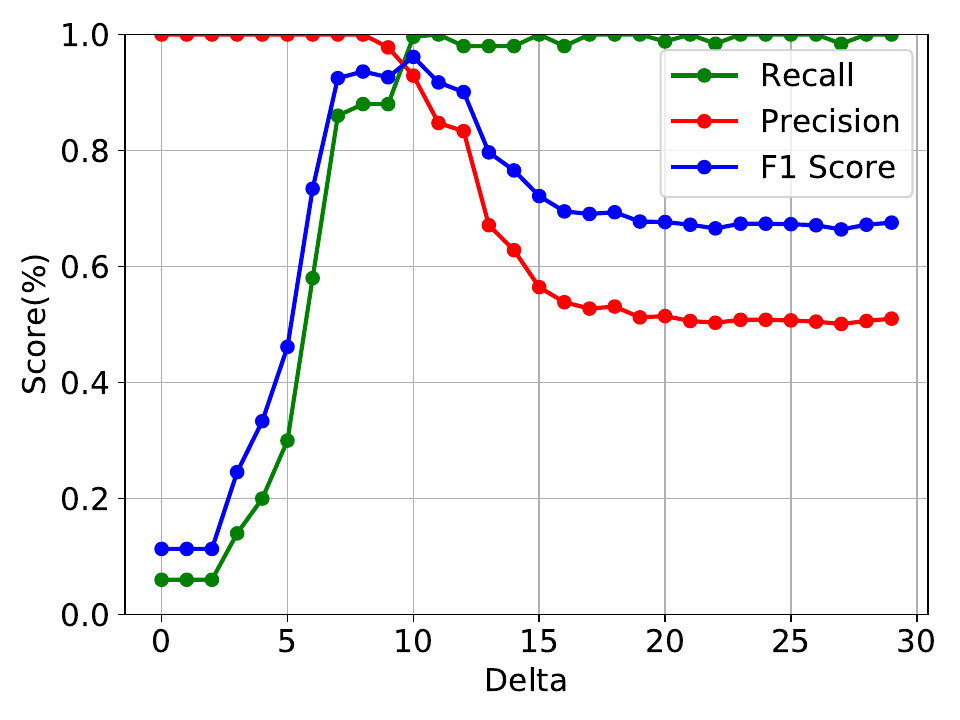}
		\caption{\textcolor{black}{
		A further illustration of the results of repeated six-fold cross validation: mean of three performance measures (recall, precision and F1 score)
		plotted as a function of the half-width $\Delta$ of the uncertainty set. 
Here, we have used a low constant mean $\mu = 5$ and high standard deviation $\sigma = 6.0$, which correspond to PSNR = 22.26 dB. }
		}
		\label{fig:var_allinone}
\end{figure}

\section{Experimental evaluation}
\label{sec:exp}


To evaluate our approach, we have implemented our \textcolor{black}{matrix-factorization} component in Apache Spark~\cite{Zaharia2010spark}, in order to ensure its scalability, and the \textcolor{black}{subspace proximity tester} in Python, using Numpy~\cite{stefan2011numpy} for numerical linear algebra and {\small \texttt{multiprocess}} for 
parallel processing.
The experiments were executed on a standard PC equipped with an Intel i7-7820X  CPU and 64~GB of RAM. 
Having said that, the execution of the sub-space proximity tester is certainly possible in many embedded systems currently available. 

\subsection{The Data}
\label{sec:data}

To validate the ability of our proposed method to detect events, we evaluated it on both synthetic and real-world datasets. Considering the limitations of the benchmarks in the literature \cite{wu2020current}, we used data from traffic monitoring collected by the Sydney Coordinated Adaptive Traffic System (SCATS) system of Dublin City Council (DCC) from intersections in Dublin, Ireland, between January 1 and November 30, 2017. 
Therein, each time series is obtained by one induction loop at an approach to an intersection with sensors at stop-lines and irregular intervals from stop-lines.
Overall, our data contains readings from $3432$ such sensors, distributed across the city.
To use a realistic data set, reflecting the asynchronous operations of the system,
we record the samples as they arrive asynchronously and do not impute
any missing values.
In particular, each intersection operates asynchronously, with all predefined phases changing, in turn, within a cycle time varying between 50 and 120 seconds both across the intersections and over time. 
Whenever an intersection's cycle time finishes, 
we record the flow over the cycle time.
Within any given period, e.g., 2 minutes, we receive vehicle count data from only a fraction of the sensors. For each day, we consider data between 7 a.m. and 10 p.m.,
which are of particular interest to traffic operators. 

Altogether, the data from $3,432$ sensors recorded with sampling period of $2$ minutes, or shorter, are flattened to a \textcolor{black}{partial} matrix $X\in \mathbb{R}^{304 \times 299,430}$, where there were $38,767,895$ zeros out of the $91,026,720$ elements, representing $42\%$ sparsity. 
This is due to a large part to the asynchronicity of the sensor readings,
and to a lesser part due to actual sensor failures.
To evaluate our approach, we have created several matrices from $X$: matrix $Y$ with a small amount of noise, which represents normal behaviour, and matrix $G$ with additional noise, which represents events.
\textcolor{black}{We have repeated this process in a repeated six-fold cross validation (out-of-sample testing) and we report the mean and standard deviation of the performance measures across the six runs.}

\textcolor{black}{In each run, using rows of matrix $X$, we have created matrices $Y$ and $Y'$ in the following way.
First, we have constructed the matrix $Y\in \mathbb{R}^{1,200 \times 299,430}$ 
representing normal behaviour in several steps. 
In the first step, each row of $Y$ has been initialised with one row sampled uniformly at random (with repetition) from \textcolor{black}{the 304 rows of} matrix $X$. 
In the second step, 
we have multiplied each row with a random scalar sampled (independently) from the uniform distribution over $(0, 2)$.
In the third step, we have applied a perturbation by an independently identically uniformly distributed noise on $[- 0.8  \Delta , 0.8  \Delta]$.
Thus constructed matrix $Y$ represents 1,200 time series of normal behaviour.
Next, we have introduced events $Y$, obtaining $Y' \in \mathbb{R}^{1,200 \times 299,430}$, or rather five variants thereof. 
In particular, we have sampled $200$ rows of matrix $Y$  uniformly at random to create $G$, which is a $200 \times 299,430$ submatrix of $Y$.
From $G$, we have created five variants of $G' \in \mathbb{R}^{200 \times 299,430}$ by the addition of Gaussian noise with mean $\mu=5,10,15,20,25$ and standard deviation equal to one half of the mean. 
This corresponds to the peak signal-to-noise ratio (PSNR) of 
23.58, 19.97, 16.94, 14.71, and 12.84 dB, respectively, when averaged over the six runs, where PSNR is the ratio between the maximum possible power of a signal and the power of the corrupting noise introduced, that is \begin{align*}
\textrm{PSNR}(G,G') = 20 \log_{10}(\max_{ij}(\{G_{ij}\})/\alpha(G,G'))\end{align*} for root mean square error \begin{align*}\alpha(G,G') = \|G' - G\|_F = \sqrt{\sum_{i=1}^{200} \sum_{j=1}^{299,430} (G'_{ij} - G_{ij})^2},\end{align*} where $G$ is the $200 \times 299,430$ submatrix of $Y$ that $G'$ is based on.
Subsequently, we have worked with a variant $Y'$ of $Y$, wherein the submatrix $G$ is replaced by $G'$.
Our training data were $1,000$ rows of this new matrix $Y'$ chosen uniformly at random and we left the remaining $200$ rows as the ground truth for testing. 
}
Using the left-out $200$ rows, we have evaluated our model with respect to recall, precision, and the so-called F1 score, which is a harmonic mean of the former two measures.


 \subsection{The Results}
 \label{sec:results}
 
 Figure~\ref{fig:iterations} \textcolor{black}{presents the performance of our  matrix-factorization component}, while Figure~\ref{fig:time_error_comparison} presents a trade-off between time required for training and reconstruction error in the choice of the rank $r$. 
 Notice that the reconstruction error is the usual extension of the root mean square error (RMSE) to the matricial setting, i.e., the Frobenius norm of the difference between the matrix $Y$ and the product $LR$.
 It is clear that increasing the rank above $10$ leads to marginal improvements in the reconstruction error, but increasing it above $40$ leads to a sharp increase in the training time. 
 We chose $r=10$ for our experiments.
 
 Figure~\ref{fig:deviation} compares the readings of sensors from the non-event  matrix $Y$ with events in $G$, while omitting zeros. We can observe that $G$ with $\mu=5$ is hard to distinguish from $Y$. 
 Figure~\ref{fig:distance_historical} presents the distribution of the values of the samples used for training: the average values of normal samples we used as input plotted in green and the average values of the samples of events (\emph{i.e.}, with the Gaussian noise for all mean values we used) plotted in red and yellow. We can observe that the supports of the distributions overlap, and especially in the case of $\mu=5$, \textcolor{black}{PSNR 23.58 dB,} the event data seem hard to distinguish from non-event data. 
 In Figure~\ref{fig:heatmap} we illustrate a heatmap of the validation matrix. The upper half of the matrix contains the normal readings while the lower half contains the event readings when $\mu=35$ is used for the noise.
 To evaluate the performance of our subspace proximity tester, we have measured recall, precision, and F1-score using different values of $\Delta$ on the $4$ matrices $G$.
 
 Figure~
 \ref{fig:syth_recall_precision_f1} presents the evolution of recall, precision, and F1-score as a function of $\Delta$ for \textcolor{black}{5} different values of $\mu$. 
 As can be observed, for small values of $\Delta$, the precision is high, while recall is low, because small values of $\Delta$ lead are more likely to
 lead to infeasibility of the LP, and hence the negative result of the test. 
 As we increase $\Delta$, we observe that our approach identifies more of the input as Normal. 
 On $G'$ matrices with $\mu$ ranging from 15 to 25, we can observe that 
 values $\Delta \in [10, 15]$ lead to the perfect performance 
 with F1-Score of $1.0$, \textcolor{black}{which should not be too surprising, considering that this regime corresponds to PSNR below 20 dB.}
 We can also observe that for noise of a lesser magnitude ($\mu=5$),
 the subspace proximity tester is able to identify samples from $G'$ with maximum F1-score \textcolor{black}{of approximately $~0.8$ for $\Delta = 8$.}
 By increasing $\Delta$ beyond this value, precision falls to $50\%$,
 which is due to the fact that too many input samples are classified as non-event.
This behaviour is to be expected, because by increasing the value of $\Delta$, we are ``relaxing'' the constrains of the linear program, which in turn leads to the Normal outcome being more common. 
 
\textcolor{black}{
Figures \ref{fig:var_syth_recall_precision_f1} and \ref{fig:var_allinone} illustrate a more challenging scenario: we keep the mean of the Gaussian noise low at $\mu = 5$, but vary the standard deviation $\sigma = 1, 2, \ldots, 6$. This corresponds to peak signal-to-noise ratios (PSNR) within 22--24 dB, as detailed in the legend of the figures. Just as above, there is a setting of $\Delta = 10$, where the F1 score approaches 1.0.  
}
 
 \textcolor{black}{Last but not least}, we note that in order to classify a new sample in dimension $\mathbb{R}^{(1\times 299,430)}$, our subspace proximity tester requires approximately $\sim0.009$ seconds for subset of cardinality $s=\log{r} \log{(r/e)}$ to obtain $e=0.1$.
 We note that this does not use the algorithm of \cite{Clarkson1987},
 and hence can be improved by many orders of magnitude, if needed.

\subsection{Benefits of Online Optimization}

Next, to demonstrate the benefits of the pursuit in the time-varying setting, we conducted the following experiment. 
We took 200 rows from our matrix $X$ of the previous section, which corresponded to normal readings from the sensors. 
Then, for $i = 0 \ldots 200$, we sampled a row from $X$,  
and added noise, which had mean $\mu = 5$ and zero variance. 
The 200 rows thus added represented slowly increasing traffic
volumes, which we would like the algorithm to adapt to.

We compare three variants of the algorithm. One, which we call ``offline'', obtains the estimate of $R_{200}$ and then does not update it further, $R_{200} = R_{200 + i}, i = 0 \ldots 200$.
Another ``online'' variant performs $n r$ updates  \eqref{eq:delta_L} and $r m$ updates \eqref{eq:delta_R} (which is known as 1 epoch) between receiving rows $i$ and $i+1$, $i = 0 \ldots 199$.
(We pick $\overline\tau$ so as to have the cardinality of $\hatSr = \hatSc$ equal to the number of hardware threads.)
Finally, another online variant performs two epochs  
between receiving rows $i$ and $i+1$, $i = 0 \ldots 199$.
\textcolor{black}{We refer to \cite{akhriev2020pursuit} for a detailed discussion of such online algorithms.}
Figure~\ref{fig:retrain} presents the resulting evolution of the RMSE. 
As can be expected, using two epochs per update (and hence more CPU cycles) performs better than using one epoch per update, which in turn performs considerably better than the offline version, whose error increases over time.

\section{Related work}
\label{sec:related}

There is much related work in change-point, anomaly, outlier, and event detection, \textcolor{black}{and the related problem of attack detection \cite{pasqualetti2013attack}}.
Since the work of Lorden~\cite{lorden1971procedures,barnett1984outliers},
there has been much work on change-point detection in univariate time series.
See~\cite{lorden1971procedures,barnett1984outliers,basseville1993detection} for a book-length history and \cite{8926446} for an overview of the latest developments.
Within anomaly detection, most statistical approaches have been tested, including hypothesis testing \cite{7428828}, dimension reduction \cite{8048463}, variants of filtering \cite{9039599}, and Gaussian processes \cite{7428828}.
In Computer Science, Complex Event Processing \cite{artikis2014heterogeneous,7000542,7024105} and deep-learning methods  \cite{9146846,8794857,9197677,9112195} are popular. 
Within change-point detection \textcolor{black}{\cite{li2019system,gu2020request}, such as cumulative statistics thresholding (CUSUM) or adaptive online thresholding (AOT),} there are relatively few papers on the multi-variate problem~\cite[e.g.]{aston2014change,cho2015multiple,zou2015efficient,Wang2018},
and fewer still, which allow for missing data~\cite{xie2013change,Soh2015}.
\textcolor{black}{Some of the recent ones \cite{feng2013online,6376091,6497613,7553443,8847621,balzano2018streaming} also consider low-rank factorizations, albeit without subsampling.} 
From the methodological works, we differ in our assumptions (uniform, rather than Gaussian noise),
focus on efficient algorithms (subsampled subspace proximity testers) for the test, and our PAC guarantees.

Our approach builds upon a rich history of research in low-rank matrix completion methodologically.
There, \cite{fazel2002matrix} suggested to replace the rank with the nuclear norm in the 
objective.
The corresponding use of semidefinite programming (SDP) has been very successful in theory \cite{candes2009exact}, 
while augmented-Lagrangian methods \cite{jaggi2010simple,Lee2010,shalev2011large,wang2014rank} and 
alternating least-squares (ALS) algorithms \cite{srebro2004maximum,Rennie2005}
have been widely used in practice  
\cite{srebro2004maximum,Rennie2005,mnih2007probabilistic,4470228,4803763}.
As it turns out \cite{keshavan2010matrix,Jain2013},
they also allow for perfect recovery in some settings.
The inequality-constrained variant of matrix completion, which we employ, 
has been introduced by \cite{marecek2017matrix}
and extended towards on-line applications in Computer Vision by \cite{akhriev2020pursuit}.

IoT applications of anomaly detection are numerous and varied \cite{li2019system,li2019online,gu2020request}, mirroring much of the development in change-point, anomaly, outlier, and event detection at large. 
As suggested in the introduction, notable examples include transportation applications \cite{6376091,6497613,6763098,RSSA:RSSA12178,kong2018lotad,7553443,8004441}, power systems \cite{farajollahi2017location,7428828,7908945,8369444}, manufacturing \cite{kanawaday2017machine,9039599} and environmental applications \cite{8081731}. 
These are, clearly, only some sample references in a much larger field.

In particular, the related work to our motivating application of IoT in Urban Traffic Management goes back at least to \cite{west1969proposed}. 
More recently, \cite{6763098} proposes a method for detecting traffic events 
that have an impact on the road traffic conditions by extending the Bayesian Robust Principal Component Analysis. They create a sparse structure composed of multiple traffic data streams (\emph{e.g.}, traffic flow and road occupancy) and use it to localize traffic events in space and time. The data streams are subsequently processed so that with little computational cost they are able to detect events in an on-line and real-time fashion.
\cite{RSSA:RSSA12178} analyze road traffic accidents based on their severity using a space-time multivariate Bayesian model. They include both multivariate spatially structured and unstructured effects, as well a temporal component  to capture the
dependencies between the severity and time effects within a Bayesian hierarchical formulation. 

Beyond the Internet of Things, Computer Vision studies a large number of related problems within ``background modelling'', where the aim is to distinguish moving objects from stationary or dynamic backgrounds in a video feed.
These are closely related to event detection, although typically focus on a single video feed, uniformly sampled, with no missing data.
We refer to the recent handbook \cite{bouwmans2016handbook}  
and to the August 2018 special issue of the Proceedings of the IEEE \cite{8425660} for up-to-date surveys.
Compared to the work in Computer Vision, we develop both subsampled subspace proximity testers (point-to-subspace distance queries), and focus on the needs of applications in IoT, where there is more variety of less reliable data sources. 



%
\section{Conclusions}

    Within a framework for representing what is an event and what is a non-event considering heterogeneous data, which are possibly not sampled uniformly, with missing values and measurement errors,
  we have presented a novel randomized event detection technique, implemented via a point-to-subspace distance query, with guarantees within probably approximately correct (PAC) learning. This is the first time such guarantees have been provided for any subsampling in matrix completion. The proofs use elaborate techniques from computational geometry (a bound on the VC dimension). 
We have also presented an experimental evaluation on data from a traffic-control system in Dublin, Ireland, which shows that it is possible to process data collected from thousands of sensors over the course of one year within minutes, 
to answer point-to-subspace distance queries in milliseconds and thus detect even hard-to-detect events. 
    We envision that this approach may have wide-ranging applications,
    wherever asynchronous high-dimensional data streams need to be monitored.

\small 
\vspace{\baselineskip}
\par\noindent
\parbox[t]{\linewidth}{
\noindent {\bf Jakub Mare{\v c}ek }\
received 
his PhD
degree from the University of Nottingham, Nottingham, U.K., in 2012.
Currently, he is a faculty member at the Czech Technical University in Prague, the Czech Republic.
He has also worked in two start-ups, at ARM Ltd., at the University of Edinburgh, at the University of Toronto, 
at IBM Research -- Ireland, 
and at the University of California, Los Angeles.
His research interests include the design and analysis of algorithms for optimisation and control problems across a range of application domains.
}
\vspace{0.5\baselineskip}
\par\noindent
\parbox[t]{\linewidth}{
\noindent {\bf Stathis Maroulis }\
received his undergraduate degree in Computer Science at the University of Athens
and his MSc from the Athens University of Economics and Business. He is currently a Ph.D. student in the Computer Science Department of the Athens University of Economics and Business under the supervision of Vana Kalogeraki.
}
\vspace{0.5\baselineskip}
\par\noindent
\parbox[t]{\linewidth}{
\noindent {\bf Vana Kalogeraki}\ 
received her PhD degree from the University of California, Santa Barbara in 2000. She holds an M.S. and a B.S. from the University of Crete, Greece. 
She is a professor at the Department of Informatics, Athens University of Economics and Business and a director of the Computer Systems and Telecommunications Laboratory. She has been working in the field of distributed and real-time systems, distributed stream processing, resource management, and fault tolerance for more than 20 years and has published more than 180 journal and conference papers and contributions to books. She has received several awards for her work (including an ERC Starting Independent Researcher Grant). 
}
\vspace{0.5\baselineskip}
\par\noindent
\parbox[t]{\linewidth}{
\noindent {\bf Dimitrios Gunopulos }\
received his PhD degree from Princeton University in 1995. Currently, he is a professor in the Department of Informatics and Telecommunications, University of Athens, Greece. He has held positions as a postdoctoral fellow at MPII, Germany, a research associate with IBM Research in Almaden, and assistant, associate, and full professor of computer science and engineering at the University of California,  Riverside. His research interests include data mining, knowledge discovery in databases, databases, sensor networks, and peer-to-peer systems. 
}

\bibliographystyle{IEEEtran}
\bibliography{completion} 

\appendix
\section{Proof of the Main Theorem}
\normalsize

As suggested earlier, our goal is to prove Theorem \ref{ourThm}, which we restate here for convenience:


    \begin{thm*}
    (i) When the distance \eqref{eq:infty} is $D \le \Delta$, Algorithm \ref{alg:SPC} never reports the point is outside the sub-space.
    (ii) When the distance \eqref{eq:infty} is $D > \Delta$,
    because there are $\epsilon n$ coordinates $i$ such that for all $\hat c$, there is $|x_i - (\hat c R)_i| \ge \Delta$,
    then for any $\delta \in (0, 1)$,
    when Algorithm \ref{alg:SPC} 
    considers $s$ coordinates 
    $$O\left( \frac{1}{\epsilon} \log \frac{1}{\delta} + \frac{r \log r}{\epsilon} \log \frac{r \log r}{\epsilon} \right)$$
    sampled independently uniformly at random, the point is inside the subspace with probability $1 - \delta$.
    \end{thm*}
    
    To see (i), consider the linear program constructed in Algorithm \ref{alg:SPC} and 
    notice that its constraints are a subset of those in \eqref{eq:inftyLP}.
    If \eqref{eq:inftyLP} is feasible, then any subset of constraints will be feasible.

    To see (ii), we show that a set related to the polyhedron of feasible $x$ has a small Vapnik-Chervonenkis (VC) dimension and apply classical results from discrete geometry.
    In particular, we proceed in four steps:
    \begin{enumerate}
        \item denote by $\mathcal{S}_1$ the range space for all possible constraints added in Line \ref{constraint1} and by $\mathcal{S}_2$ the range space for all possible constraints added in Line
    \ref{constraint2}. 
    \item The VC dimension of each of $\mathcal{S}_1, \mathcal{S}_2$ is at most $r + 1$.
    \item The VC dimension of $\mathcal{S}_1 \cup \mathcal{S}_2$ is $O(r \log r)$.
    \item Subsequently, we apply the celebrated result:
        \end{enumerate}
    
    \begin{thm}[\cite{Haussler1986,Haussler1987,Clarkson1986,Clarkson1987}]
     \label{Haussler}
    Let $(\mathcal{X},\mathcal{R})$ be a range space of Vapnik-Chervonenkis dimension $d$. Let $\epsilon, \delta \in (0, 1)$.
    If $\mathcal{S}$ is a set of $$O\left( \frac{1}{\epsilon} \log \frac{1}{\delta} + \frac{d}{\epsilon} \log \frac{d}{\epsilon} \right)$$ points 
    sampled independently from a finite subset of $\mathcal{X}$, then $\mathcal{S}$ is an $\epsilon$-net for the finite subset with probability at least 
    $1 - \delta$.
    \end{thm}    
    
To develop these ideas formally, let us reiterate the usual definitions of discrete geometry using the notation of \cite{vapnik1971uniform} and \cite{Haussler1987}, which partly overlaps with the notation used in the paper. We use calligraphic fonts in this appendix to distinguish S of the main body of the paper from $\mathcal{S}$ of the appendix, etc. 

\begin{dfn}[Range space of \cite{vapnik1971uniform}]
A range space $\mathcal{S}$ is a pair
$(\mathcal{X},\mathcal{R})$, where $\mathcal{X}$ is a set and $\mathcal{R}$ is a family of
subsets of $\mathcal{X}$, $\mathcal{R} \subseteq 2^{\mathcal{X}}$. Members of $\mathcal{X}$ are called
elements or points of $\mathcal{S}$ and members of
$R$ are called ranges of $\mathcal{S}$. $\mathcal{S}$ is finite if
$\mathcal{X}$ is finite.
\end{dfn}

Notice that the range space is a (possibly infinite) hypergraph.

\begin{dfn}[Shattering of \cite{vapnik1971uniform}]
 Let $\mathcal{S} = (\mathcal{X},\mathcal{R})$ be a range
space and let $\mathcal{A} \subset \mathcal{X}$ be a finite set. Then $\Pi_\mathcal{R}(\mathcal{A})$ denotes the set
of all subsets of $\mathcal{A}$ that can be obtained
by intersecting $\mathcal{A}$ with a range of $\mathcal{S}$.
If $\Pi_{\mathcal{R}}(\mathcal{A}) = 2^{\mathcal{A}}$, we say that $\mathcal{A}$ is shattered
by $\mathcal{R}$. 
\end{dfn}

\begin{dfn}[Dimension of \cite{vapnik1971uniform}]
The Vapnik-Chervonenkis
dimension of $\mathcal{S}$ is the smallest integer $d$ such that no $\mathcal{A} \subset \mathcal{X}$ of cardinality $d + 1$ is shattered
by $\mathcal{R}$. If no such $d$ exists, we say the
dimension of $\mathcal{S}$ is infinite.
\end{dfn}

\begin{dfn}[$\epsilon$-net of \cite{Haussler1986}]
An $\epsilon$-net of a finite subset of points $P \subseteq \mathcal{X}$ is a subset $\mathcal{N} \subseteq P$ such that any range $\mathcal{r} \in \mathcal{R}$ with $|\mathcal{r} \cap P| \ge \epsilon |P|$ has a non-empty intersection with  $\mathcal{N}$.
\end{dfn}

\paragraph{Step 1.} The range spaces $\mathcal{S}_1$ and $\mathcal{S}_2$ will 
share the same set of points, namely $[n] := {1, 2, \ldots,  n}$,
and feature very similar ranges:  $\mathcal{S}_1$ will feature the
hyperplanes 
$ x_i - (\hat c R)_i \le \Delta $
corresponding to the first set of constraints in the LP \eqref{eq:inftyLP},
while $\mathcal{S}_2$ will feature the
hyperplanes 
$(\hat c R)_i - x_i \le \Delta$.
We keep them separate, so as to allow for the hyperplanes to be
in a generic position.

Alternatively, one could construct a single range space, with the
same set of points and ranges given by the subspaces given by the intersections of $ x_i - (\hat c R)_i \le \Delta $
and 
$(\hat c R)_i - x_i \le \Delta$
for $i \in [n]$.
This would, however, complicate the analysis, somewhat. 

\paragraph{Step 2.} 
The VC dimension of each of $\mathcal{S}_1, \mathcal{S}_2$ is at most $r + 1$.
For range spaces, where the ranges are hyper-planes, this is a standard result. We refer to Section 15.5.1 of  \cite{Gaertner2012} for a very elegant proof using Radon's theorem. 
Notice that $r$ would suffice, if there were no vertical hyperplanes.

\paragraph{Step 3.}
The VC dimension of $\mathcal{S}_1 \cup \mathcal{S}_2 $ is $O(r \log r)$.
This follows by the counting of the possible ranges and Sauer-Shelah lemma, a standard result. We refer to Lemma 15.6 in  \cite{Gaertner2012}.

\paragraph{Step 4.}
The intuition is that if there is a large-enough subset,
a large-enough random sample will intersect with it.
The surprising part of Theorem~\ref{Haussler} on the existence of $\epsilon$-nets is that the bound of the large-enough
does not depend on the number of points of the ground
set, but only on the VC dimension established above.
In particular, 
we sample coordinates $S, |S| = s$ in Line \ref{line:sample}. This corresponds to sampling from $\mathcal{X}$ in
$\mathcal{S}_1 \cup \mathcal{S}_2$.
Because we assume there are $\epsilon n$ coordinates $i$
such that such that for all $\hat c$, there is $|x_i - (\hat c R)_i| \ge \Delta$, an $\epsilon$-net will intersect these by Theorem~\ref{Haussler}.

\end{document}